\documentclass{ws-ijmpe}
\usepackage[super,compress]{cite}
\begin{document}

\markboth{Mamta Aggarwal \& G. Saxena}{Persistence of magicity in neutron rich exotic $^{78}$Ni in ground as well as excited states}

\catchline{}{}{}{}{}
%%%%%%%%%%%%%%%%%%%%%%%%%%%%%%%%%%%%%%%%%%%%%%%%%%%%%%%%%%%%%%%%%%%%

\title{PERSISTENCE OF MAGICITY IN NEUTRON RICH EXOTIC $^{78}$Ni IN GROUND AS WELL AS EXCITED STATES}

\author{Mamta Aggarwal}

\address{Department of Physics, University of Mumbai, Kalina Campus, Mumbai-400098, India\\
mamta.a4@gmail.com}

\author{G. Saxena}

\address{Department of Physics, Govt. Women Engineering College, Ajmer-305002, India\\
gauravphy@gmail.com}

\maketitle

\begin{history}
\received{Day Month Year}
\revised{Day Month Year}
%\accepted{Day Month Year}
%\comby{(xxxxxxxxxx)}
\end{history}

\begin{abstract}
Recent experimental observation of magicity in $^{78}$Ni has infused the interest to examine the persistence of the magic character across the N$=$50 shell gap in extremely neutron rich exotic nucleus $^{78}$Ni in ground as well as excited states. A systematic study of Ni isotopes and N$=$50 isotones in ground state is performed within the microscopic framework of relativistic mean-field (RMF) and the triaxially deformed Nilson Strutinsky model (NSM). Ground state density distributions, charge form factors, radii, separation energies, pairing energies, single particle energies and the shell corrections show strong magicity in $^{78}$Ni. Excited nuclei are treated within the statistical theory of hot rotating nuclei where the variation of level density parameter and entropy shows significant magicity with a deep minima at N$=$50, which, persists up to the temperatures $\approx$ 1.5$-$2 MeV and then slowly disappear with increasing temperature. Rotational states are evaluated and effect of rotation on N$=$50 (Z$=$20$-$30) isotones are studied. Our results agree very well with the available experimental data and few other theoretical calculations.\par
\end{abstract}


\ccode{PACS numbers: 23.50.+z, 21.10.-k, 21.10.Dr}

%\tableofcontents

\section{Introduction}
The phenomenal advances in the nuclear experimental techniques \cite{tanihata,tanihata1,ozawa,ozawa1} during the last two decades have made it possible to study more and more exotic nuclei. One such recent experimental observation of magicity in $^{78}$Ni has invoked tremendous curiosity to explore the unique characteristics of such extremely neutron rich exotic doubly magic nucleus. The persistence of the magic character across N$=$50 shell gap in neutron rich nuclei has attracted a lot of attention of nuclear theorists to test their models in this yet unknown regime. As the very neutron rich nuclei play an important role in the r-process nucleosynthesis, the doubly magic $^{78}$Ni is of special relevance in the astrophysical r-process path as it serves as one of the possible waiting points in the synthesis of heavier elements \cite{hosmer}. Among the classic nuclear shell gaps and doubly magic nuclei, only $^{48}$Ni, $^{78}$Ni, $^{100}$Sn, and $^{132}$Sn are are close to nucleon drip lines away from the stability. Out of these, $^{78}$Ni is the most exotic one and with experimentally unknown properties. Therefore, $^{78}$Ni, with the extreme N/Z ratio in a doubly magic nucleus and the proximity to the neutron dripline and the continuum, represents a unique stepping stone towards the physics of extremely neutron-rich nuclei.\par

Several experimental and theoretical efforts are going on to study the Ni and its neighboring isotopes to examine the magicity away from the stable valley and validate many experimental and theoretical investigation with their properties. The conventional magic numbers Z(N)$=$2, 8, 20, 28, 50, 82, 126 have so far continued to play an important role in the theoretical and experimental development of nuclear physics. However, the break down of conventional magicity N$=$8, 20, 28 etc. \cite{iwasaki,door,bastin} and confirmation of new magicity N$=$14, 16, 32, 34 \cite{stanoiu,brown,becheva,kanungo,hoffman,tshoo, gade,wienholtz,rosenbusch,stepp} along with Z$=$16 \cite{togano1} have provided routes of consistent improvement of various theoretical calculations and experimental techniques. Specifically, in the case of $^{78}$Ni, the large neutrons to protons ratio might allow the observation of unusual shell effects. For shell-model calculations where this nucleus is considered as a core \cite{sieja,sieja1,sieja2,urban} in order to explore the properties of A$=$80$-$90 neutron-rich nuclei, including those lying on the r-process path, the knowledge of single-particle energies of $^{78}$Ni would be crucial to know. \par

In the last decade, the half life of $^{78}$Ni has been deduced experimentally for the first time in Refs $~$\cite{hosmer,schatz,xu}. A systematic study of $\beta$ decay half lives has provided the experimental indication of double magicity \cite{xu} for both the Z$=$28 and N$=$50 shell gaps in $^{78}$Ni. In-beam $\gamma$-ray spectroscopy of low-lying level structures of nuclei in the vicinity of $^{78}$Ni also indicated its doubly magicity \cite{shiga}. Recently, two back to back studies by Olivier \textit{et al.} \cite{olivier} and  Welker \textit{et al.} \cite{welker} have hallmarked $^{78}$Ni as a strong doubly magic nucleus. The doubly magic character in $^{78}$Ni has already been reported by many theoretical calculations including shell-model calculations \cite{sieja}, relativistic mean-field model using TMA parameter \cite{yadav1}, relativistic continuum Hartree-Bogoliubov theory \cite{meng} and First-Principles computations \cite{hagen}. Furthermore, the sensitivity to the symmetry energy and the role of the continuum in neutron-rich Ni isotopes has been discussed by Piekarewicz \textit{et al.} \cite{piekarewicz}. Shell evolution above Z, N$=$50 within Skyrme density functional theory has been analyzed with a blocking procedure which accounts for the polarization effects, including deformations \cite{shi}. In addition, the signature of fifth island of inversion, i.e., a well-deformed prolate band at low excitation energy in $^{78}$Ni, providing a striking example of shape coexistence far from stability \cite{nowacki}. \par

In a very recent work, the temperature dependence of the symmetry energy and neutron skins in Ni, Sn, and Pb isotopic chains \cite{antonov} of magic nuclei have indicated the importance of the study of the excited states of the neutron rich doubly magic nuclei. The first 2$^+$ state in $^{78}$Ni is predicted at nearly 4 MeV, a value analogous to the 2$^+$ state of $^{132}$Sn \cite{sieja}. An observation of an 8$^+$ isomeric state in $^{78}$Zn for the first time, represents the closest approach to the doubly-magic nucleus $^{78}$Ni made so far in $\gamma$-spectroscopy studies \cite{daugas}. This result consists of the first experimental evidence of the persistence of the N$=$50 shell gap near $^{78}$Ni. Since the shell structure gets influenced profoundly by the excitations, it is important to understand the behaviour of N$=$50 shell gap in neutron rich nuclei in the vicinity of  $^{78}$Ni due to the excitation energy at different temperatures and spins, which, along with a systematic study of ground state properties, is the objective of this work. \par

In this paper, we study the persistence of the magic character of $^{78}$Ni in two parts:\\
(1) We study the ground state of $^{78}$Ni within the theoretical framework of RMF+BCS approach ~\cite{yadav,saxena,saxena1,saxena2} and the triaxially deformed Nilson Strutinsky Model (NSM) \cite{aggarwal,MAMPRC}, and compare our results with existing experimental and other theoretical data.\\
(2) The effect of temperature and spin on doubly magic exotic nucleus $^{78}$Ni and its neighbouring N$=$50 isotones (Z$=$20$-$30) is introduced for hot rotating nuclei similar to that done in our earlier works \cite{aggarwal,MAPLB}. The excited states are treated using the statistical theory of hot rotating nuclei \cite{aggarwal,MAPLB,mamijmpe1,mamijmpe2}. Rotational spectrum at various temperatures along with the statistical properties such as the level density parameter (LD) \cite{bethe,ericson} and entropy for N$=$50 shell gap in the vicinity of $^{78}$Ni, are studied.\par

\section{Ground state properties}

\subsection{RMF approach}

Here, we present a detailed investigation of ground state properties of the entire chain of Ni isotopes with N$=$20$-$70 and N$=$50 isotones for Z$=$20$-$50. For this purpose, we employ relativistic mean-field plus state dependent BCS (RMF+BCS) approach \cite{walecka,boguta,suga,ring,yadav,saxena,saxena1,saxena2} together with a realistic mean-field, which has proved to be very useful and a successful tool especially for drip line nuclei as shown in our earlier work \cite{saxena,saxena1,saxena2}. We use the model Lagrangian density with nonlinear terms both for the ${\sigma}$ and ${\omega}$ mesons as described in detail in Refs.~\cite{suga,yadav,singh}. For the pairing interaction, we use a delta force, i.e. V$=$-V$_0 \delta(r)$ with the strength V$_0$$=$350 MeV fm$^3$ which has been used in Refs.~\cite{yadav,saxena,saxena1,saxena2} for the description of drip-line nuclei. Based on the single-particle spectrum calculated by the RMF, we perform a state dependent BCS calculations \cite{lane,ring2}. The detailed description of the theoretical formalisms that have been adequately described in our earlier works have not been given in this paper. (Readers may refer ~ \cite{suga,yadav,singh,saxena1,meng5} for detailed description of formalism). \par

\begin{figure}[htb]
\centering
\includegraphics[width=0.8\textwidth]{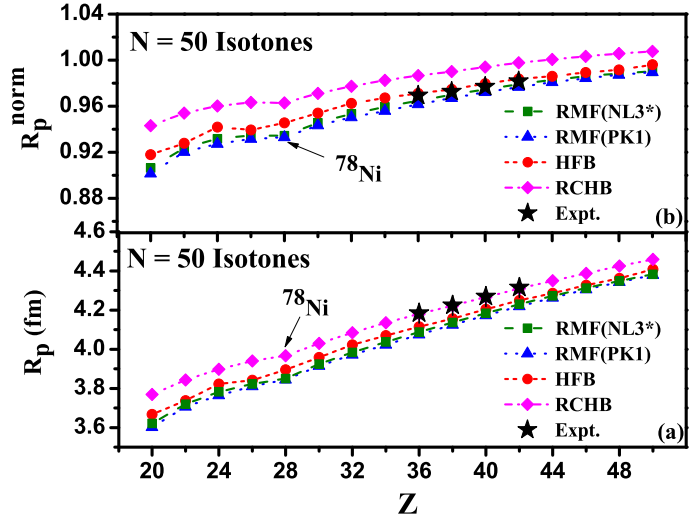}
\caption{(Colour online) (a) Variation of proton rms radii with respect to Z for N$=$50 isotones. (b) normalized radii (R$_{p}^{nor}$$=$R$_{p}$/R$_{p}^{Co}$) vs. Z.}
\label{fig1}
\end{figure}

There have been various attempts to identify the emergence of the non-traditional magic numbers \cite{angeli1,angeli2} based on the analysis of the systematics of the experimental proton radii reported recently. Therefore, the size of a nucleus, which can be defined as the root-mean-square (rms) radius of its nucleon distribution, is expected to provide important insights on the evolution of the magic numbers \cite{tran,kumawat}. In view of this, we have plotted proton rms radii for N$=$50 isotones (Z$=$20$-$50) calculated by RMF approach using NL3* \cite{nl3star} and PK1 \cite{pk1} parameters in Fig. 1(a). For a comparison, we have also plotted available experimental proton radii extracted from Ref.$~$\cite{angeli1} using R$_{p}$$=$$\surd$R$_{c}^2$-0.64 fm relation. In addition, we have shown the proton radii calculated by other theories like non-relativistic approach viz. Skyrme-Hartree-Fock method with the HFB-24 functional \cite{goriely} and relativistic continuum Hartree-Bogoliubov (RCHB) theory with the relativistic density functional PC-PK1 \cite{xia}. Angeli \textit{et al.} \cite{angeli1,angeli2} have suggested a kink (change in the slope) for shell closure \cite{angeli1,angeli2} which is evident in Fig. 1(a) for the case of $^{78}$Ni from all the theories considered. To eliminate the smooth mass number dependence of the proton rms radii R$_{p}$, we normalize proton rms radii using the following formula given by Collard \textit{et al.} \cite{collard}
\begin{equation}
R_{p}^{Co} = \sqrt{3/5}\,\, (1.15 + 1.80A^{-2/3} - 1.20A^{-4/3})\,\,A^{1/3} fm
\end{equation}
The normalized radii (R$_{p}^{nor}$$=$R$_{p}$/R$_{p}^{Co}$) is also plotted in Fig. 1(b) from all the theories considered and also with the available experimental data. One can see a sharp kink once again showing the magicity of $^{78}$Ni.\par

\begin{figure}[htb]
\centering
\includegraphics[width=0.8\textwidth]{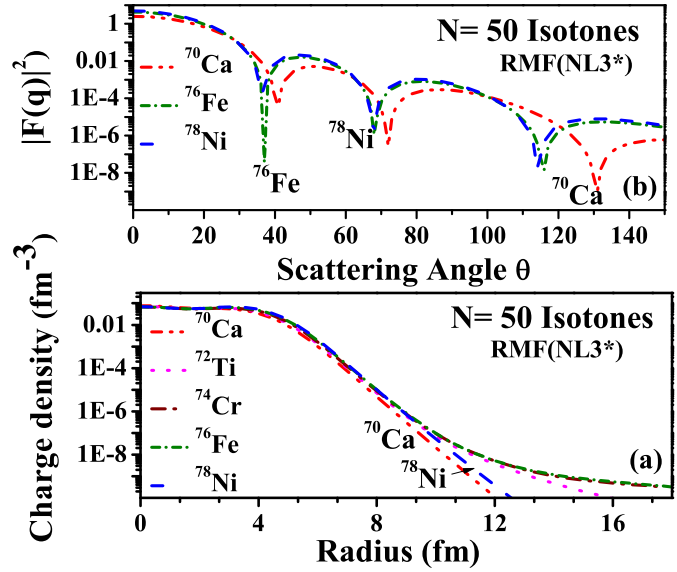}
\caption{(Colour online) (a) Variation of charge distribution along with radius (b) Variation of Form Factor with respect to scattering angle for some selected nuclei in N$=$50 Isotones.}
\label{fig2}
\end{figure}

The charge distribution for various isotones of N$=$50 including that of $^{78}$Ni is estimated. The radial charge distribution of $^{70}$Ca, $^{72}$Ti, $^{74}$Cr, $^{76}$Fe and $^{78}$Ni (Z ranging from 20 to 28) is displayed in Fig. 2(a). Interestingly, the charge distribution of $^{78}$Ni with the largest Z ($=$28) among the nuclei considered, is found similar to the charge distribution of another magic number with lowest Z ($=$ 20). The charge densities of $^{78}$Ni and $^{70}$Ca are found to fall rapidly as compared to that of $^{72}$Ti, $^{74}$Cr and $^{76}$Fe. This pattern of density, which is confined to smaller distances, further exhibits magicity in $^{78}$Ni. \par

A useful physical observable, the nuclear charge form factor, a measurable quantity through the elastic electron-nucleus scattering experiments \cite{hofstadter,forest,donnelly} is displayed in Fig. 2(b) for $^{70}$Ca, $^{76}$Fe and $^{78}$Ni. A clear difference in the peaks of form factor of $^{76}$Fe and $^{78}$Ni can be easily visible around scattering angle 40$^{\circ}$. The difference between form factor of $^{70}$Ca and $^{78}$Ni at large scattering angle $\sim$130$^{\circ}$ may be attributed  to the halo structure of $^{70}$Ca which is reported by Meng \textit{et al.} \cite{meng02} and Kaushik \textit{et al.} \cite{kaushik}. However, near future projects like SCRIT~\cite{suda,suda1} and ELISe~\cite{antonov,simon} are expected to provide much needed data to identify magic nature experimentally by measurement of charge distribution.

\begin{figure}[htb]
\centering
\includegraphics[width=0.8\textwidth]{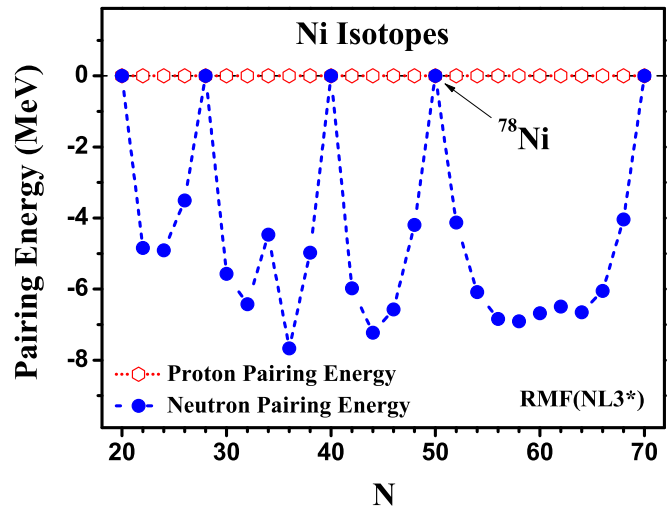}
\caption{(Colour online) Pairing energy contribution from protons and neutrons for Ni isotopes.}
\label{fig3}
\end{figure}

The pairing energy contribution from the protons and neutrons has been computed for Ni isotopes using NL3* parameter, which has been displayed in Fig. 3. Proton pairing energy has always been found zero characterizing magic character at Z$=$28. Neutron pairing energy varies from zero to 8 MeV in between two magic numbers therefore zero contribution from neutron side along with zero proton pairing energy lead to the nuclei which are doubly magic. Therefore, we get $^{48,56,68,78,98}$Ni as doubly magic candidates as seen in Fig. 3. Similar behaviour is found from PK1 parameter (not shown here).

\begin{figure}[htb]
\centering
\includegraphics[width=0.8\textwidth]{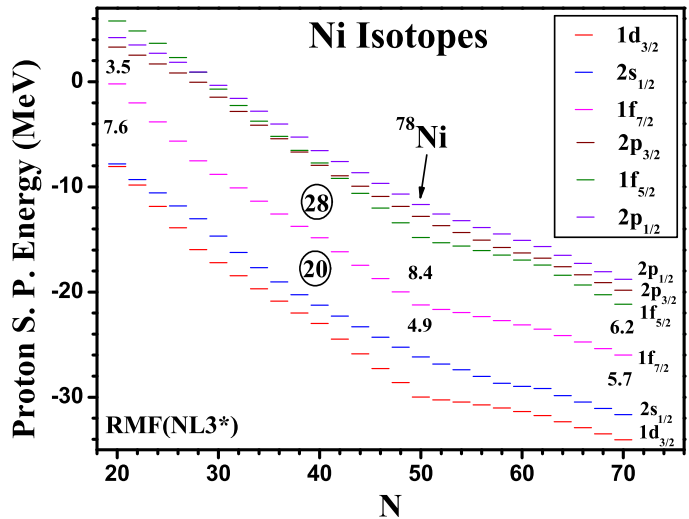}
\caption{(Colour online) Proton single particle energies for Ni isotopes. Gaps (in MeV) between few states for selected nuclei are shown by numbers.}
\label{fig4}
\end{figure}

The proton single particle energies for Ni isotopes are plotted in Fig. 4, which show the variation of the proton single particle states with the neutron number (N$=$20$-$70) variation. The gap between the proton s-d shell and proton 1f$_{7/2}$ state is significant enough and leads to Z$=$20 shell closure. The gap between 1f$_{7/2}$ and proton fp shell (1f$_{5/2}$, 2p$_{3/2}$ and 2p$_{1/2}$) indicate shell closure at Z$=$28. Both these gaps are sufficient enough to provide proton shell closures for all the neutron numbers shown. However, for N$=$50, the shell gap of Z$=$20 decreases gradually and the shell gap responsible for Z$=$28 attains its maximum value of 8.4 MeV (at 7$^{78}$Ni) which builds up from 3.5 MeV (at $^{48}$Ni) to 6.2 MeV (at $^{98}$Ni). This shows that at $^{78}$Ni, the proton single particle states are aligned to produce strongest magicity.\par

\begin{figure}[htb]
\centering
\includegraphics[width=0.8\textwidth]{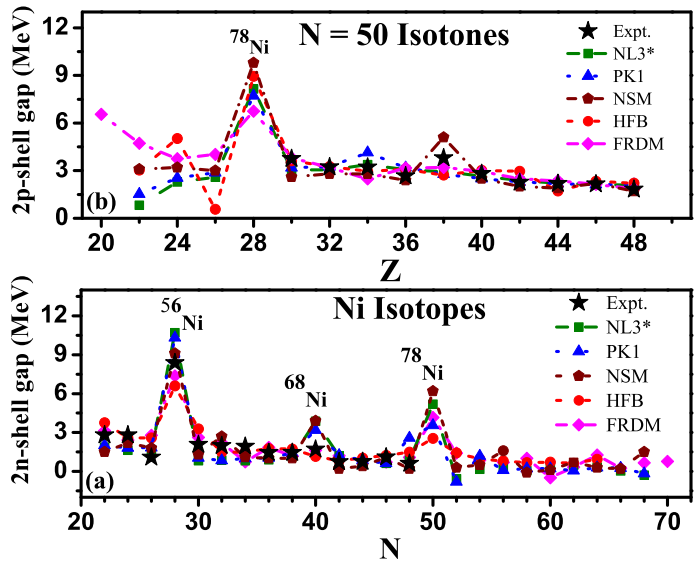}
\caption{(Colour online) Two neutron shell gap [S$_{2n}$(N, Z)$-$S$_{2n}$(N+2, Z)] and two proton shell gap [S$_{2p}$(N, Z)$-$S$_{2p}$(N, Z+2)] for Ni isotopes and N$=$50 isotones in (a) and (b), respectively.}
\label{fig5}
\end{figure}

In Fig. 5(a) and (b), we compare two neutron shell gap [S$_{2n}$(N, Z)$-$S$_{2n}$(N+2, Z)] and two proton shell gap [S$_{2p}$(N, Z)$-$S$_{2p}$(N, Z+2)] for Ni isotopes and N$=$50 isotones, respectively. Here, we plot our results of RMF using NL3* \cite{nl3star} and PK1 parameters \cite{pk1} along with the results using Nilson Strutinsky Model (NSM) \cite{aggarwal,MAMPRC} (given in next section) for comparison, in Fig. 5. In addition, results of other theories like non-relativistic approach viz. Skyrme-Hartree-Fock method with the HFB-24 functional \cite{goriely} and Finite Range Droplet Model (FRDM) \cite{frdm2012} are also plotted along with available experimental data \cite{wang}. Fig. 5 itself speaks its excellent match with experiments, both parameter of RMF, NSM approach and other theories. Here the peak refers to magic character which are apparent in Figs. 5(a) and (b) for the case of $^{78}$Ni.

\subsection{Nilson Strutinsky Model (NSM)}

\begin{figure}[htb]
\centering
\includegraphics[width=0.8\textwidth]{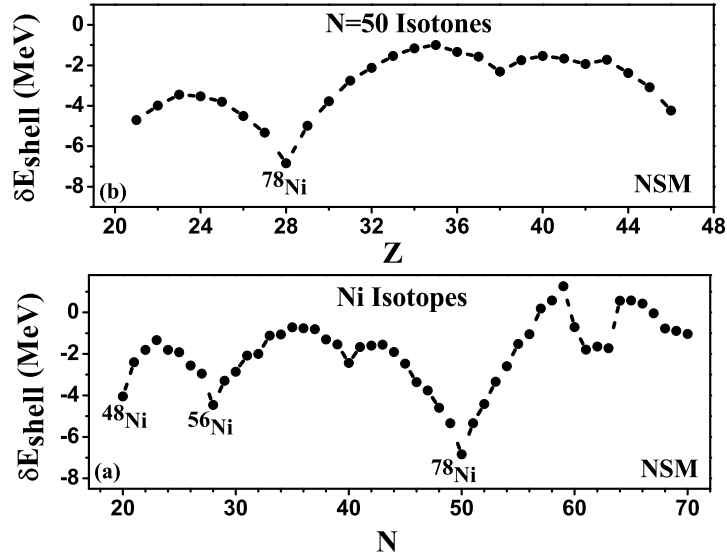}
\caption{Values of shell correction $\delta$E$_{Shell}$ to energy (in MeV) vs. neutron number N for Ni isotopes and vs. proton number Z for N$=$50 isotones in (a) and (b), respectively.}
\label{fig6}
\end{figure}

Another formalism used for the present study is the triaxially deformed Nilson Strutinsky (NS) model which treats the delicate interplay of macroscopic bulk properties of nuclear matter and the microscopic shell effects, and has been used extensively in our earlier works ~\cite{aggarwal,MAMPRC}. We evaluate binding energy, separation energy, deformation and shape by incorporating macroscopic binding energy BE$_{LDM}$ obtained from the LDM mass formula ~\cite{PM} to the microscopic effects arising due to nonuniform distribution of nucleons through the Strutinsky's shell correction $\delta$E$_{shell}$ ~\cite{VM} along with the deformation energy E$_{def}$ obtained from the surface and Coloumb effects ~\cite{MAMPRC}. Energy E ($=$-BE) minima are searched for various $\beta$ (0 to 0.4 in steps of 0.01) and $\gamma$ (from -180$^o$ (oblate) to -120$^o$ (prolate) and -180$^o$  $<$ $\gamma$ $<$ -120$^o$ (triaxial)) to trace the nuclear shapes and equilibrium deformations.

\begin{figure}[htb]
\centering
\includegraphics[width=0.9\textwidth]{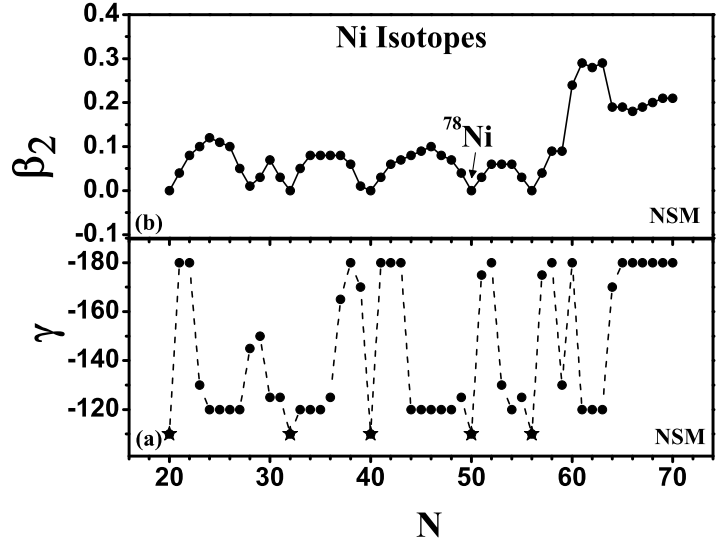}
\caption{(a) Shape parameter $\gamma$ and (b) deformation($\beta$) vs N for Ni isotopes. In $\gamma$ plot, asterisk symbols represent spherical nuclei.}
\label{fig7}
\end{figure}

Fig. 6 shows very strong evidence for the magic character of $^{78}$Ni in the plots of shell correction to energy. The shell correction $\delta$E$_{Shell}$, which is expected to show peaks at around shell closures, is found to show deepest minima at $^{78}$Ni, both for Ni isotopes and N$=$50 isotones, indicating a very strong magic character. The sphericity and the magic character of other magic nuclei (at N$=$20 and 28 in Fig. 6(a)) also show minima showing magicity. N$=$40 show small peak as well indicating semi magic character. $\delta$E$_{Shell}$ values plotted in Fig. 6, vary from a minimum of around $\sim$ 4-5 MeV near closed shells as expected to a few keV's around other nuclei. At $^{78}$Ni, the  $\delta$E$_{Shell}$ values is around 8 MeV which is the deepest minima among all isotopes of Ni.

\begin{figure}[htb]
\centering
\includegraphics[width=0.9\textwidth]{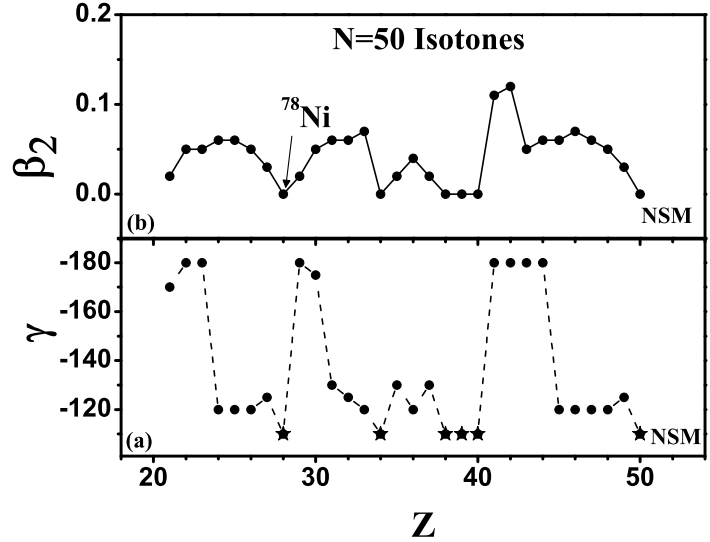}
\caption{ Shape parameter $\gamma$ and (b) deformation($\beta$) vs Z for N$=$50 isotones. In $\gamma$ plot, asterisk symbols represent spherical nuclei.}
\label{fig8}
\end{figure}

The shape parameter $\gamma$ and equilibrium quadrupole deformation $\beta$ of Ni isotopes are plotted in Figs. 7(a)  and 7(b), respectively. Nuclei with double magic nucleons with N$=$20, 28, 40(semi magic), 50  are spherical with zero deformation. Mid-shell nuclei show deformation upto 0.1 although they are magic with Z$=$28. Shapes change rapidly from oblate to triaxial to prolate and spherical. Fig. 8 shows beta and gamma for N$=$50 isotones where deformation of all isotones are usually small and the shape is spherical, oblate or triaxial.

\section{Excited nuclei- Statistical theory of hot rotating nuclei}

Excited high spin states of the neutron rich exotic nuclei across N$=$50 shell gap are studied within the framework of statistical theory of hot rotating nuclei ~\cite{aggarwal,MAPLB,mamijmpe1,mamijmpe2,MAJNPT}. Since the shell structure, which is mainly responsible for the equilibrium deformation or shape of the nucleus in the ground state of a nucleus, eventually gets wash out by heating of the nucleus at a certain critical temperature, , we assess the persistence of magicity of N=50 in and near $^{78}$Ni with increasing excitation. We study statistical properties at different temperatures (T) and spin (M($\hbar$)). For this purpose, we use statistical theory of hot rotating nuclei and evaluate excitation energy (E$_x$) and entropy (S) of the system. We minimize the free energy (F$=$E$-$TS) to trace deformation and shape of the nuclei at angular momentum M$=$0$-$60$\hbar$. (Kindly refer to Refs.  ~\cite{aggarwal,MAPLB} for detailed description of theoretical formalism).

This section has two parts where\\
(i) first, we suppress the rotational degree of freedom and study the temperature dependence of N$=$50 shell closure, and then\\
(ii) the influence of rotation across N$=$50 shell gap is examined. \par

\subsection{Temperature effects}
\begin{figure}[htb]
\centering
\includegraphics[width=0.7\textwidth]{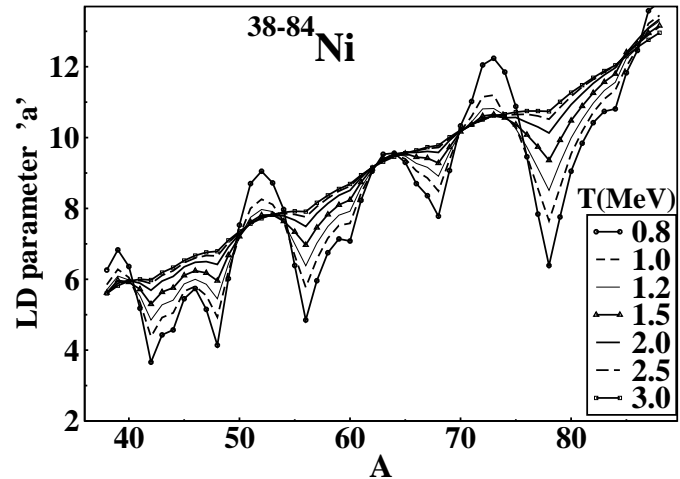}
\caption{Level density parameter vs. mass number A for Ni isotopes at different temperatures.}
\label{fig9}
\end{figure}

It is now a well-known fact that the density of the quantum mechanical states increases rapidly with excitation energy and the nucleus shifts from discreteness to quasi-continuum to continuum where the statistical concepts like level density (LD)~\cite{bethe,ericson,MAKAILAS,MALD,MAJNPLD}, which is the number of excited levels around an excitation energy, become crucial for the prediction of various nuclear and astrophysical phenomena. Since the level density parameter 'a' is expected to be minimum at the shell closures, we evaluate and plot the level density parameter 'a' ($=$S$^2$/4E$_x$) of Ni isotopes at different temperatures in Fig. 9. We use  a wide range of Ni isotopes $^{46-84}$Ni which includes doubly magic $^{48}$Ni, $^{56}$Ni and exotic nucleus $^{78}$Ni.  Deep minima at N$=$20, 28, and 50 shows the strong magic character upto the temperatures (T) around 1.5$-$2 MeV. With further increasing T, shell effects appear to diminish and 'a' variation becomes more and more smooth with N and the kink in the curve disappears completely at T$=$3 MeV. Deepest minima at $^{78}$Ni shows strong evidence of magicity which has persisted upto a high temperature of T$=$1.5$-$2 MeV. \par

\begin{figure}[htb]
\centering
\includegraphics[width=0.7\textwidth]{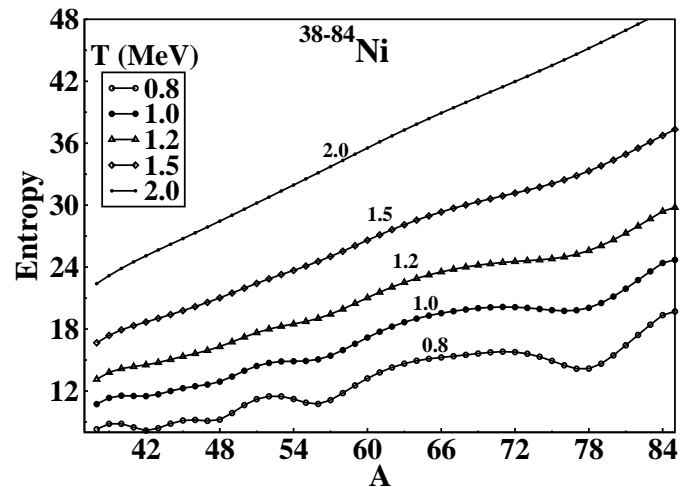}
\caption{Entropy vs. mass number A for Ni isotopes at different temperatures.}
\label{fig10}
\end{figure}

Entropy, which always increases with increasing excitation, is shown for the Ni isotopes at different T$=$0.8$-$2.0 MeV in Fig. 10. One may see a minima in the entropy at N$=$20, 28 and 50 which is an evidence of magicity. Entropy is constantly increasing with isospin as well as T, but shows a minima at shell closures especially at low T$=$0.8-1.5 MeV. $^{78}$Ni shows a peak with the lowest value as compared to N$=$20 and 28 shell closures. This proves the strong magic character of $^{78}$Ni. For higher temperatures $>$ 1.5 MeV, entropy increases gradually. Disappearance of shell structure with increasing excitation is evident in level density parameter as well as entropy plots of neutron rich $^{78}$Ni which establishes this nucleus as strongly doubly magic. \par

\begin{figure}[htb]
\centering
\includegraphics[width=0.7\textwidth]{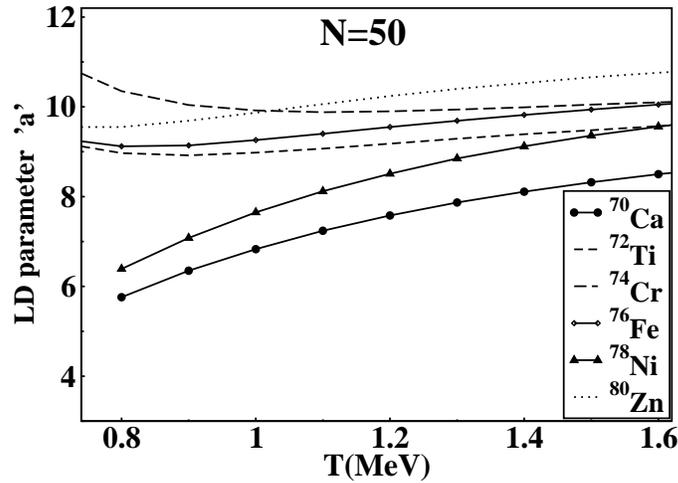}
\caption{Level density parameter 'a' vs. T for N$=$50 (Z$=$20$-$30) isotones.}
\label{fig11}
\end{figure}

The level density parameter of N$=$50 isotones (Z$=$20$-$30) are plotted in Fig. 11. It may be noted that the doubly magic neutron rich $^{70}$Ca and $^{78}$Ni show the lowest level density parameter among all the other N$=$50 isotones which reaffirms the doubly magic characteristic of exotic $^{78}$Ni as well as $^{70}$Ca (to be shown in detail in our upcoming works).\par

\subsection{Rotation effects}

\begin{figure}[htb]
\centering
\includegraphics[width=0.8\textwidth]{Fig.12.eps}
\caption{Rotational states for spin values from 0 to 60 $\hbar$ (a) Deformation ($\beta$) and (b) E$_{rot}$ vs. M($\hbar$) for $^{78}$Ni at different temperatures(T).}
\label{fig12}
\end{figure}

\begin{figure}[htb]
\centering
\includegraphics[width=0.8\textwidth]{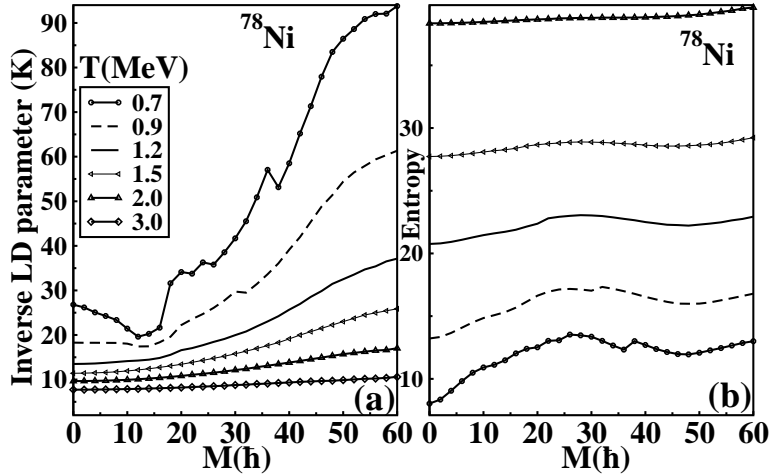}
\caption{(a) Inverse level density parameter 'K'$=$A/a and (b) Entropy vs. M($\hbar$) for $^{78}$Ni at different temperatures(T). At low T, the effect of rotation and shell structure are predominant.}
\label{fig13}
\end{figure}

The structural properties of any nucleus are profoundly altered once the rotational degree of freedom is included.  At low T, the shell structure as well as the effects of rotation are predominant. The equilibrium state of the nucleus is strongly influenced by the changes in the single particle shell structure due to the temperature and spin. To assess the impact of rotation on the strong magic character in the vicinity of $^{78}$Ni, we incorporate rotational degree of freedom and evaluate excitation energy which includes rotational energy (E$_{rot}$) as well thermal excitation energy.

Fig. 12 (a) shows the deformation ($\beta$) as a function of angular momentum at T$=$0.7$-$3.0 MeV. It is interesting to see the doubly magic nucleus getting deformed with rotation. At zero spin, $\beta$ is zero which increases upto a value 0.16 with unusual prolate non-collective shape, which eventually undergoes a shape transition to the usual oblate non-collective shape phase with deformation increasing upto 0.2. As T increases, the deformed shape phase at low spin disappears and the nucleus directly goes from spherical (with zero deformation) to oblate shape phase which is classical response. At very high T, the nuclear deformation at high spin is much higher than at low T. Fig. 12(b) shows rotational energy  of $^{78}$Ni as a function of M. We note a kink at low T$=$0.7 MeV which disappears with increasing T. \par

\begin{figure}[htb]
\centering
\includegraphics[width=0.8\textwidth]{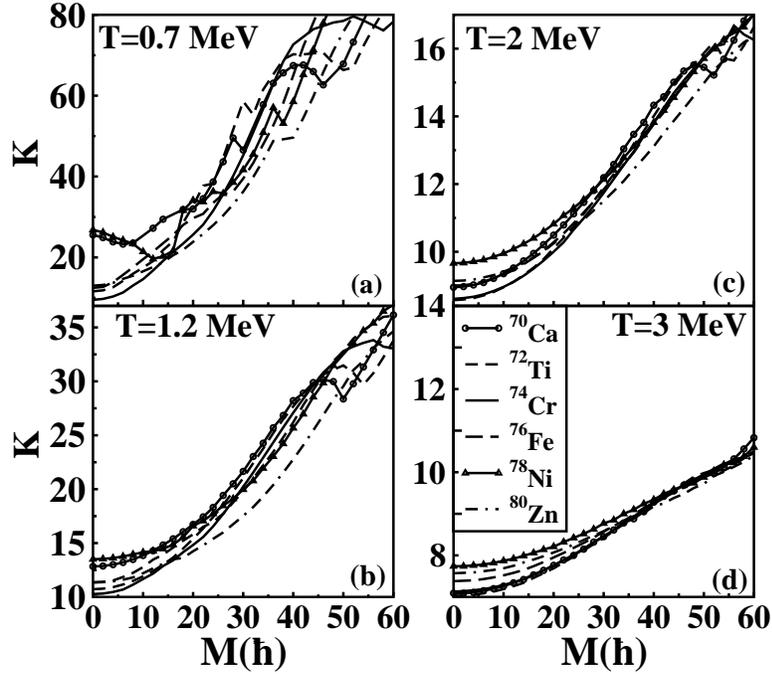}
\caption{'K' vs. M($\hbar$) at different temperatures for N$=$50 (Z$=$20$-$30). At T$=$3 MeV, K is almost constant around 9$-$10 as expected.}
\label{fig14}
\end{figure}

Fig. 13 shows (a) inverse level density parameter K$=$(A/a), and (b) entropy of $^{78}$Ni varying with spin for different T which show fluctuations at low T due to varying shell structure as shown in our earlier work ~\cite{MALD}. With increasing T, shell effects are washed out and 'K' attains almost constant value $\approx$ 9$-$10 as expected. The entropy varies smoothly for higher T. \par

\begin{figure}[htb]
\centering
\includegraphics[width=0.8\textwidth]{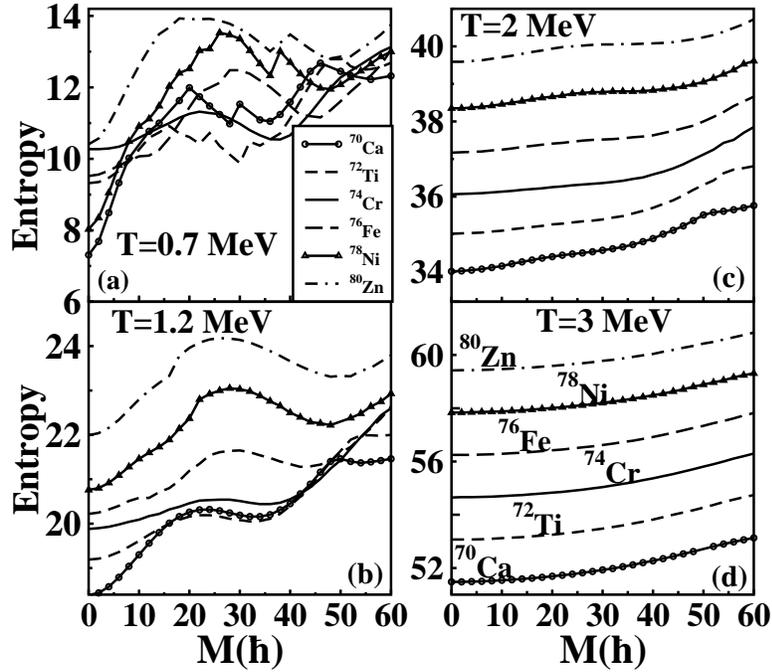}
\caption{Entropy vs. M($\hbar$) for N$=$50 isotones (Z$=$20$-$30) at different temperature. Almost constant entropy at high T shows washing away of shell structure.}
\label{fig15}
\end{figure}

Inverse level density parameter 'K' and entropy for N$=$50 isotones (Z$=$20$-$30) close to $^{78}$Ni are plotted in Figs. 14 and 15, respectively. At low T, we see fluctuations in 'K' as a function of M which gets smoothed out with increasing T. 'K' for $^{78}$Ni is the highest for all T. At T$=$3 MeV where 'K' becomes almost constant at a value around 9$-$10, Ni remains the highest among all the isotones proving its magic character at high excitations.\par

\section{Summary}
A detailed investigation of magicity in the ground and excited states of $^{78}$Ni and N$=$50 isotones is performed within the microscopic frameworks. Our calculations include single particle spectra, proton and neutron densities, charge form factors, radii and other ground state properties calculated by NL3* and PK1 force parameters of RMF. A detailed analysis of radii and density distribution of Ni isotopes and N$=$50 isotones bear witness of strongest magic character of $^{78}$Ni. Also, pairing energy contribution, single particle levels, two-neutron (proton) shell gap calculated by RMF and the deformation, shape and shell correction ($\delta$E$_{Shell}$) to energy calculated by Nilson Strutinsky model reveal strong magic character of $^{78}$Ni. Our calculations are in close match with experimental data and the other theories viz. HFB, RCHB and FRDM. \par

Excited states of $^{78}$Ni and N$=$50 isotones for Z$=$20$-$30 are studied as a function of temperature and spin. Level density parameter and entropy show minima at N$=$50 shell gap showing strong magicity in $^{78}$Ni. Among N$=$50 isotones (Z$=$20$-$30), LD parameter is minimum for doubly magic neutron rich $^{78}$Ni and $^{70}$Ca.\par

Rotational states of $^{78}$Ni are computed. Rotation degree of freedom induces deformation in $^{78}$Ni which undergoes a shape transition from an unusual prolate non-collective at low spins to oblate non-collective at high spins. At high T, the nucleus remains spherical at low spins with a transition to oblate at high spins which is the usual behaviour of any hot rotating nucleus. Inverse level density parameter for $^{78}$Ni is maximum among the other N$=$50 isotones showing the magic character of this exotic neutron rich nucleus.\par

\section*{Acknowledgements}
Authors would like to thank Prof. H. L. Yadav, BHU, India and Prof. L. S. Geng, Beihang University, China for their kind guidance and support. Authors G. Saxena and M. Aggarwal gratefully acknowledge the support provided by Science and Engineering Research Board (DST), Govt. of India under the young scientist project YSS/2015/000952 and WOS-A scheme respectively.


\begin{thebibliography}{0}
\bibitem{tanihata} I. Tanihata {\it et al.}, P\textit{hys. Lett. B} {\bf 206} (1988) 592.
\bibitem{tanihata1} Rituparna Kanungo, I. Tanihata and A. Ozawa, \textit{Phys. Lett. B} {\bf B 512} (2001) 261.
\bibitem{ozawa} A. Ozawa, T. Kobayashi, T. Suzuki, K. Yoshida and I. Tanihata,  \textit{Phys. Rev. Lett.} {\bf 84} (2000) 5493.
\bibitem{ozawa1} A. Ozawa {\it et al.}, \textit{Nucl. Phys. A} {\bf 709} (2002) 60.
\bibitem{hosmer} P. T. Hosmer \textit{et al.}, \textit{Phys. Rev. Lett.} \textbf{94} (2005) 112501.
\bibitem{iwasaki} H. Iwasaki \textit{et al.}, \textit{Phys. Lett. B} {\bf 481} (2000) 7.
\bibitem{door} P. Doornenbal \textit{et al.}, \textit{Phys. Rev. Lett.} \textbf{111} (2013) 212502.
\bibitem{bastin} B. Bastin \textit{et al.}, \textit{Phys. Rev. Lett.} {\bf 99} (2007) 022503.
\bibitem{stanoiu} M. Stanoiu \textit{et al.}, \textit{Phys. Rev. C}  \textbf{69} (2004) 034312.
\bibitem{brown} B. Alex Brown and W. A. Richter, \textit{Phys. Rev. C} {\bf 72} (2005) 057301.
\bibitem{becheva} E. Becheva \textit{et al.}, \textit{Phys. Rev. Lett.} \textbf{96} (2006) 012501.
\bibitem{kanungo} R. Kanungo \textit{et al.}, \textit{Phys. Lett. B} \textbf{528} (2002) 58.
\bibitem{hoffman} C. R. Hoffman \textit{et al.}, \textit{Phys. Lett. B} {\bf 672} (2009) 17.
\bibitem{tshoo} K. Tshoo \textit{et al.}, \textit{Phys. Rev. Lett.} {\bf 109} (2012) 022501.
\bibitem{gade} A. Gade \textit{et al.}, \textit{Phys. Rev. C} \textbf{74} (2006) 021302(R).
\bibitem{wienholtz} F. Wienholtz \textit{et al.}, \textit{Nature} \textbf{498} (2013) 346.
\bibitem{rosenbusch} M. Rosenbusch \textit{et al.}, \textit{Phys. Rev. Lett.} \textbf{114} (2015) 202501.
\bibitem{stepp} D. Steppenbeck \textit{et al.}, \textit{Nature} \textbf{502} (2013) 207.
\bibitem{togano1} Y. Togano \textit{et al.}, \textit{Phys. Rev. Lett.} {\bf 108} (2012) 222501.
\bibitem{sieja} K. Sieja and F. Nowacki \textit{Phys. Rev. C} \textbf{85} (2012) 051301(R).
\bibitem{sieja1} K. Sieja, F. Nowacki, K. Langanke, and G. Martinez-Pinedo, \textit{Phys. Rev. C} \textbf{79} (2009) 064310.
\bibitem{sieja2} G. S. Simpson \textit{et al.}, \textit{Phys. Rev. C} \textbf{82} (2010) 024302.
\bibitem{urban} W. Urban \textit{et al.}, \textit{Phys. Rev. C} \textbf{85} (2012) 014329.
\bibitem{schatz} H. Schatz \textit{et al.}, \textit{Eur. Phys. J. A} \textbf{25} (2005) 639.
\bibitem{xu} Z. Y. Xu., textit{Phys. Rev. Lett.} \textbf{113} (2014) 032505.
\bibitem{shiga} Y. Shiga \textit{et al.}, \textit{Phys. Rev. C} \textbf{93} (2016) 024320.
\bibitem{olivier} L. Olivier \textit{et al.}, \textit{Phys. Rev. Lett.} \textbf{119} (2017) 192501.
\bibitem{welker} A. Welker \textit{et al.}, \textit{Phys. Rev. Lett.} \textbf{119} (2017) 192502.
\bibitem{yadav1} H. L. Yadav, S. Sugimoto, and H. Toki, \textit{Mod. Phys. Lett. A} \textbf{17} (2002) 2523. 
\bibitem{meng} J. Meng, H. Toki, S. G. Zhou, S. Q. Zhang, W. H. Long and L. S. Geng, \textit{Prog. Part. Nucl. Phys.} \textbf{57} (2006) 470.
\bibitem{hagen} G. Hagen, G. R. Jansen, and T. Papenbrock, \textit{Phys. Rev. Lett.} \textbf{117} (2016) 172501.
\bibitem{piekarewicz} J. Piekarewicz, \textit{Phys. Rev. C} \textbf{91} (2015) 014303.
\bibitem{shi} Yue Shi, \textit{Phys. Rev. C} \textbf{95} (2017) 034307.
\bibitem{nowacki} F. Nowacki, A. Poves, E. Caurier, and B. Bounthong, \textit{Phys. Rev. Lett.} \textbf{117} (2016) 272501.
\bibitem{antonov} A. N. Antonov, D. N. Kadrev, M. K. Gaidarov, P. Sarriguren, and E. Moya de Guerra, \textit{Phys. Rev. C} \textbf{95} (2017) 024314.
\bibitem{daugas} J. M. Daugas \textit{et al.}, \textit{Phys. Lett. B} 476 (2000) 213.
\bibitem{yadav}  H. L. Yadav, M. Kaushik, and H. Toki, \textit{Int. Jour. Mod. Phys. E} {\bf 13} (2004) 647.
\bibitem{saxena}  G. Saxena, D. Singh, M. Kaushik, H. L. Yadav, and H. Toki, \textit{Int. Jour. Mod. Phys. E} {\bf 22} (2013) 1350025.
\bibitem{saxena1} G. Saxena, M. Kumawat, M. Kaushik, U. K. Singh, S. K Jain, S. Somorendro Singh, and Mamta Aggarwal, {Int. Jour. Mod. Phys. E} \textbf{26} (2017) 175007. \bibitem{saxena2} G. Saxena, M. Kumawat, M. Kaushik, S. K. Jain, and Mamta Aggarwal, \textit{Phys. Lett. B} \textbf{775} (2017) 126.
\bibitem{aggarwal} Mamta Aggarwal, \textit{Phys. Rev. C} {\bf 90} (2014) 064322.
\bibitem{MAMPRC} Mamta Aggarwal, \textit{Phys. Rev. C} {\bf 89} (2014) 024325 .
\bibitem{MAPLB} Mamta Aggarwal, \textit{Phys. Lett. B} {\bf 693} (2010) 489.
\bibitem{mamijmpe1} M. Aggarwal, \textit{Int. Jour. Mod. Phys. E} \textbf{17} (2008) 1091.
\bibitem{mamijmpe2} M. Rajasekaran and M. Aggarwal, \textit{Int. Jour. Mod. Phys. E} \textbf{7}(1998) 389.
\bibitem{bethe} H. Bethe, \textit{Phys. Rev.} \textbf{50} (1936) 332; \textit{Rev. Mod. Phys.} \textbf{9} (1937) 69.
\bibitem{ericson} T. Ericson, \textit{Adv. Phys.} \textbf{9} (1960) 425.
\bibitem{walecka} B. D. Serot and J. D. Walecka, \textit{Adv. Nucl. Phys.} {\bf 16} (1986) 1.
\bibitem{boguta} J. Boguta and A. R. Bodmer, \textit{Nucl. Phys. A} {\bf 292} (1977) 413.
\bibitem{suga} Y. Sugahara and H. Toki, \textit{Nucl. Phys. A} {\bf 579} (1994) 557.
\bibitem{ring}  P. Ring, \textit{Prog. Part. Nucl. Phys.} {\bf 37} (1996) 193.
\bibitem{singh} D. Singh, G. Saxena, M. Kaushik, H. L. Yadav, and H. Toki, \textit{Int. Jour. Mod. Phys. E} {\bf 21} (2012) 1250076.
\bibitem{lane} A. M. Lane, \textit{Nuclear Theory} (Benjamin, 1964).
\bibitem{ring2}  P. Ring and P. Schuck, \textit{The Nuclear many-body Problem}, Springer-Verlag (1980).
\bibitem{meng5}  J. Meng, \textit{Phys. Rev. C} {\bf 57} (1998) 1229.
\bibitem{angeli1} I. Angeli and K. P. Marinova, \textit{At. Data Nucl. Data Tables} \textbf{99} (2013) 69.
\bibitem{angeli2} I. Angeli and K. P. Marinova, \textit{J. Phys. G: Nucl. Part. Phys.} \textbf{42} (2015) 055108.
\bibitem{tran} D. T. Tran \textit{et al.}, arXiv:1709.03355v1 (2017).
\bibitem{kumawat} M. Kumawat, G. Saxena, M. Kaushik, R. Sharma and S. K. Jain, \textit{Canadian Jouranl of Physics} (2018), arXiv:1804.10833.
\bibitem{nl3star} G. A. Lalazissis, S. Karatzikos, R. Fossion, D. Pena Arteaga, A. V. Afanasjev, and P. Ring, \textit{Phys. Lett. B} {\bf 671} (2009) 36.
\bibitem{pk1} W. Long, J. Meng, N. Van Giai, and Shan-Gui Zhou, \textit{Phys. Rev. C} {\bf 69} (2004) 034319.
\bibitem{goriely} S. Goriely \textit{et al.},\textit{ Phys. Rev. Lett.} \textbf{102} (2009) 152503, http://www-astro.ulb.ac.be/bruslib.
\bibitem{xia} X. W. Xia \textit{et al.}, \textit{At. Data Nucl. Data Tables} \textbf{121-122} (2018) 1.
\bibitem{collard} H. R. Collard, L. R. B. Elton, and R. Hofstadter, \textit{Springer Berlin} \textbf{2} (1967).
\bibitem{hofstadter} R. Hofstadter, \textit{Rev. Mod. Phys.} \textbf{28} (1956) 214.
\bibitem{forest} T. de Forest Jr. and J. D. Walecka, \textit{Adv. Phys.} \textbf{15} (1966) 1.
\bibitem{donnelly} T. W. Donnelly and J. D. Walecka, \textit{Annu. Rev. Nucl. Part. Sci.} \textbf{25} (1975) 329.
\bibitem{meng02} J. Meng, H. Toki, J. Y. Zeng, S. Q. Zhang, and S.-G. Zhou, \textit{Phys. Rev. C} \textbf{65} (2002) 041302(R).
\bibitem{kaushik} M. Kaushik, D. Singh and H. L. Yadav, \textit{Acta Phys. Slov.} \textbf{55} (2005) 181.
\bibitem{suda} T. Suda and M. Wakasugi, \textit{Prog. Part. Nucl. Phys.} \textbf{55} (2005) 417.
\bibitem{suda1} T. Suda \textit{et al.}, \textit{Phys. Rev. Lett.} \textbf{102} (2009) 102501.
\bibitem{simon} H. Simon, \textit{Nucl. Phys. A} \textbf{787} (2007) 102.
\bibitem{frdm2012} P. Moller, A. J. Sierk, T. Ichikawa, and H. Sagawa, \textit{At. Data Nucl. Data Tables} {\bf109}-{\bf110} (2016) 1.
\bibitem{wang} M. Wang \textit{et al.}, \textit{Chin. Phys. C} {\bf 36} (2012) 1603.
\bibitem{PM} P. Moller, J. R. Nix, W. D. Myers, and W. J. Swiatecki, \textit{At. Data Nucl. Data Tables} {\bf{59}} (1995) 185.
\bibitem{VM} V. M. Strutinsky, Nucl. Phys. A {\bf 95} (1967) 420.
\bibitem{MAJNPT} M. Aggarwal, \textit{J. Nucl. Phy. Mat. Rad. A} \textbf{3} (2016) 179.
\bibitem{MAKAILAS} Mamta Aggarwal and S. Kailas, \textit{Phys. Rev. C} \textbf{81} (2010) 047302.
\bibitem{MALD} Mamta Aggarwal and S. kailas, \textit{Proceedings of the DAE Symp. on Nucl. Phys.} \textbf{62} (2017) 96.
\bibitem{MAJNPLD} Mamta Aggarwal, \textit{J. Nucl. Phy. Mat. Rad. A} (2018).

\end{thebibliography}
\end{document}